\documentclass[a4paper,11pt,reqno,final]{article}
\usepackage{epsfig,amssymb,amsmath,version}
\usepackage{amssymb,version,graphicx,fancybox,mathrsfs}
\usepackage[notcite,notref]{showkeys}
\usepackage{url,hyperref}
\usepackage{subfigure}
\usepackage{setspace}
\usepackage{color}
\usepackage{cases}
\usepackage{txfonts}
\usepackage{geometry}
\renewcommand  \varphi \phi

\newcommand{\bs}[1]{\boldsymbol{#1}}
\def \er {\bs  {\hat  x}}

\def \vt {{\bs T}_l^m}

\renewcommand \wedge \times

\begin{document}
\bibliographystyle{plain}
\baselineskip 12pt

\centerline{\Large\bf Fast and Accurate Computation of Exact  Nonreflecting
}\vskip 2pt \centerline{\Large\bf  Boundary Condition for Maxwell's Equations }
\vskip 20pt
\centerline{\large\bf Xiaodan Zhao  {and}  Li-Lian Wang}
\vskip 12pt
\centerline{Division of Mathematical Sciences}
\centerline{Nanyang Technological University, Singapore, 637371}
\centerline{zhao0122@e.ntu.edu.sg; \;lilian@ntu.edu.sg}
\vskip 20pt
{\noindent\bf Abstract:}  We report in this paper a fast and accurate algorithm for computing the exact spherical
nonreflecting  boundary condition (NRBC) for  time-dependent  Maxwell's equations.
 It is essentially based on a new formulation of the NRBC, which allows for the use of an  analytic method for computing the involved  inverse Laplace transform. This tool can be generically integrated with the interior solvers for challenging simulations of electromagnetic  scattering problems.  We provide some numerical examples  to show that the algorithm leads to very accurate results.

\thispagestyle{empty}
\vskip 24pt

\centerline{\bf 1. Introduction}
\vskip 12pt

The time-domain simulations, which are capable of capturing wide-band signals and modeling more
general material inhomogeneities and nonlinearities, have attracted much attention \cite{TafloveHag.05,lee1995whitney,liu1998pstd}. A longstanding issue in many simulations  resides in how to deal with the unbounded computational domain.
Various approaches including the perfectly matched layer (PML) (cf. \cite{Bere94}), the boundary integral methods (cf. \cite{CISB91}),   nonreflecting (absorbing  or transparent)  boundary conditions (cf. \cite{Kel.G89,Hagstrom99}), and many others,  have been proposed to surmount this obstacle. Although the use of exact nonreflecting boundary conditions (NRBCs) is desirable and beneficial, the practitioners are usually  plagued with their complications and  computational inefficiency.  Indeed, these time-domain NRBCs are global in both time and space, and involving Laplace inversion of special functions.
It is worthwhile to highlight some works on efficient algorithms for exact NRBCs for acoustic wave equations, see e.g.,
\cite{Sofronov1992,Alpert,Lubistian2002,LiJing2006,Wang2Zhao12}. However, there has been significantly less study of the NRBCs for  Maxwell's equations, where one  finds the existing formulations (see e.g., \cite{Hagstrom07,Chen08})
  actually  present a great challenge for evaluation.

  In this paper, we reformulate the NRBC for the three-dimensional Maxwell's equations, and extend the techniques for the NRBC of the acoustic wave equation in \cite{Alpert,Wang2Zhao12} for computing it in a fast and accurate manner. It is important to
  point out that it is quite generic to integrate this sort of semi-analytic tool with any solver for the interior truncated problem (for example, the finite element/spectral element methods, and the boundary perturbation technique \cite{Nich.Shen2009}),
  with the aid of the Spherepack \cite{adams1999spherepack} or certain hybrid mesh interpolation \cite{lin2009hybrid}.

Typically,  we consider an electromagnetic scattering problem with a homogeneous background transmission medium, and
 with a bounded scatterer $D.$    Assume that the source current (or excitation source),  other inputs and inhomogeneity of media are supported in a ball of radius $b,$ that is,  $B:=\{\bs x\in {\mathbb R}^3 : |\bs x|<b\}.$  Then the analytic method of Laplace transform and separation of variables can be applied to solve the time-dependent Maxwell's system (exterior to $B$)  with free source, homogeneous initial data and  the Silver-Muller radiation condition
\begin{equation}\label{SMradiationcond}
\partial_t \bs E_T+c\,
\bs  {\hat x}\times \partial_t \bs H
=o(|\bs x|^{-1}),\quad  t>0;\quad c=1/{\sqrt{\varepsilon\mu}},
\end{equation}
where  $\{\bs E,\bs H\}$ are the electric and magnetic fields, $\bs  {\hat  x}=\bs x/|\bs x|,$ and $\bs E_T=\bs
{\hat  x}\times (\bs E\times \bs  {\hat  x})$ is the tangential
component of $\bs E.$   The electric permittivity $\varepsilon$ and magnetic permeability  $\mu$ are
 positive constants. The underlying solution (cf. Hagstrom and Lau \cite{Hagstrom07})  can be expressed in terms of  vector spherical harmonic functions (VSHs) with the coefficients determined by the electric field  on $B$:
  \begin{equation}\label{Eexpansion}
 {\bs E}=\sum_{l=1}^\infty\sum_{m=-l}^l\Big(E_{lm}^r Y_l^m \er+E_{lm}^{(1)}\nabla_SY_{l}^m+E_{lm}^{(2)}\vt\Big), \quad
{\rm at}\;\; r=b,
 \end{equation}
where the VSHs  $\big\{ Y_l^m \er, \nabla_S Y_{l}^m, \vt:=\nabla_SY_{l}^m\times\er\big\}$ are the orthogonal basis of $(L^2(S))^3$ with $S$ being the unit sphere  (see e.g., \cite{Morse53}), and $\big\{Y_l^m\big\}$ being the spherical harmonics as normalized in  \cite{Nedelec}.  Note that in  \cite{Hagstrom07}, the exact NRBC is expressed as a system of  $\bs E$ and $\bs H,$ which is actually equivalent to the formulation (cf. \cite{Chen08}) by using the VSH notation  here:
\begin{equation}\label{NRBCold}
\partial_t \bs E_T-c\,\bs  {\hat  x}\times (\nabla \wedge \bs E)={\mathcal T}_b[{\bs E}],\quad {\rm at}\;\; r=b,
\end{equation}
where the electric-to-magnetic (EtM) operator:
\begin{equation}\label{oldEtM}
{\mathcal T}_b[\bs E]=\frac c b \sum_{l=1}^\infty\sum_{m=-l}^l\Big(
\big(\rho_l\ast {E}_{lm}^{(1)}\big)\,{\nabla_S
Y}_l^m+\big(\sigma_l\ast {E}_{lm}^{(2)}\big)\,\vt\Big).
\end{equation}
Here, $\rho_l$ and $\sigma_l$   (termed as the nonreflecting boundary kernels (NRBKs)) are defined by
\begin{equation}\label{NRBK}
\rho_l(t)={\mathcal L}^{-1}\bigg[z\,\bigg(\frac{z\, k_l(z)}{k_l(z)+z\, k_l'(z)}+1\bigg) \bigg](t), \;\;
\sigma_l(t) ={\mathcal L}^{-1}\bigg[1+z+z \frac{k_l'(z)}{k_l(z)}\bigg](t)\;\; {\rm with}\;\;  z=\frac{sb}{c},
\end{equation}
where ${\mathcal L}^{-1}$ is the inverse Laplace
transform (in $s$-domain), and $k_l(z)$ is the modified spherical Bessel function defined by $k_{l}(z)=\sqrt{2/ ({\pi z})} K_{l+1/2}(z)$, with  $K_{l+1/2}$ being the modified Bessel function of the second kind of order $l+1/2$ {\rm(}cf. \cite{watson}{\rm)}.  The involved convolution is defined as usual: $(f\ast g)(t)=\int_0^t f(\tau)g(t- \tau)d\tau.$

Now, the central task is to compute the NRBC in \eqref{NRBCold}-\eqref{NRBK}. Notice  that the electric field $\bs E$
at  $r=b$ is unknown as the NRBC serves as the boundary condition for the interior problem.  Here, we resort to the Spherepack
\cite{adams1999spherepack} to communicate between the electric field and the VSH expansion coefficients. Thus, some hybrid mesh interpolation technique (cf. \cite{lin2009hybrid})  is necessary if the spatial discretization of the interior solver (e.g., the finite/spectral element methods)  uses a different set of  grids on the sphere. Thus, the critical issue becomes how to compute the NRBKs in \eqref{NRBK}, and temporal convolutions in \eqref{oldEtM} at any time $t$ efficiently.  This will be the topic of the following section.

 \vskip 12pt
\centerline{\bf 2.~ The Algorithm for Computing the  NRBC}
  \vskip 12pt

The NRBK $\sigma_l$  appears in the NRBC for the transient wave equation, which has an explicit formula (see \eqref{kernela} below) derived from the Residue theory  (see e.g., \cite{Hagstrom07,Wang2Zhao12}). However, this analytic tool for inverse Laplace transform can not be applied to
compute $\rho_l,$  since we lack information on the zeros of $k_l(z)+zk'_l(z)$ (i.e., the poles of the integrand in the inverse Laplace transform), while that of $k_l(z)$ is available.  In fact, there is no stable way to directly compute the NRBK $\rho_l.$

   \vskip 4pt
\noindent{\bf A.~ Alternative formulation of ${\mathcal T}_b[\bs E].$}~
  \vskip 2pt

  Observe from \eqref{oldEtM} that the EtM operator only involves the VSH expansion coefficients
   $\big\{{E}_{lm}^{(1)},{E}_{lm}^{(2)}\big\}$  in \eqref{Eexpansion}. In fact, there holds the following relation between ${E}_{lm}^{r}$ and
   ${E}_{lm}^{(1)}:$
\begin{equation}\label{conffs}
\widehat E^r_{lm}(s)=l(l+1)\frac{ k_l(z)}{k_l(z)+z\, k_l'(z)}
\widehat E_{lm}^{(1)}(s),\quad z=\frac{sb}{c}, \quad {\rm at}\;\; r=b,
\end{equation}
where $\big\{\widehat E^r_{lm}(s), \widehat E^{(1)}_{lm}(s)\big\}$ are Laplace transforms of $\big\{{E}_{lm}^{r},{E}_{lm}^{(1)}\big\},$ respectively. The derivation of \eqref{conffs} is quite involved, so we will provide the proof in the extended paper.
      This leads to the following alternative formulation, from which the efficient algorithm stems. \\[2pt]

\noindent{\bf Theorem 1.} {\em  The EtM operator $\mathcal T_b[\bs E]$ in \eqref{oldEtM} can be reformulated as:
 \begin{equation}\label{newEtM}
{\mathcal T}_b[\bs E]=\frac c b \sum_{l=1}^\infty\sum_{m=-l}^l\bigg(
\frac{\omega_l\ast {E}_{lm}^{r}}{l(l+1)}\,{\nabla_S
Y}_l^m+\big(\sigma_l \ast {E}_{lm}^{(2)}\big)\,\vt\bigg),
\end{equation}
where  $\sigma_l$ is defined in \eqref{NRBK}, and $\omega_l$ is given by
\begin{equation}\label{NRBKnew}
\omega_l(t)={\mathcal L}^{-1}\bigg[z\,\bigg(1+z+z
\frac{k_l'(z)}{k_l(z)}\bigg)\bigg](t)=\frac b c\big(\sigma_l'(t)+\sigma_l(0)\delta(t)\big),\quad z=\frac{sb}{c},
\end{equation}
with $\delta$ being the Dirac delta function.}

\vskip 4pt
\noindent{\bf B.~ Explicit formulas of the NRBKs $\sigma_l$ and $\omega_l$.}~
  \vskip 4pt

As already mentioned,  the NRBK $\sigma_l$ appears in the exact NRBC for the wave equation, and
the explicit formula derived from  the Residue theory  (see e.g., \cite{Hagstrom07,Wang2Zhao12}) is
 of the form:
\begin{equation}\label{kernela}
\sigma_{l}(t)=\frac{c} b \sum_{j=1}^{l}z_j^l e^{\frac{c}{b} z_j^lt},\quad l\ge 1,\;\; t\ge 0,
 \end{equation}
where $\{z_j^l\}_{j=1}^l$ are the zeros of  $K_{l+1/2}(z).$ Hence, it follows from  \eqref{NRBKnew} that
\begin{equation}\label{kernel}
 \omega_l(t)=\frac c b \sum_{j=1}^l (z_j^l)^2 e^{\frac{c}{b} z_j^lt}+\delta(t)\sum_{j=1}^l z_j^l,  \quad l\ge 1,\;\; t\ge 0.
 \end{equation}

\vskip 2pt

\begin{figure}[!ht]
\subfigure{
\begin{minipage}[t]{0.44\textwidth}
\centering
\rotatebox[origin=cc]{-0}{\includegraphics[width=0.75\textwidth,height=0.7\textwidth]{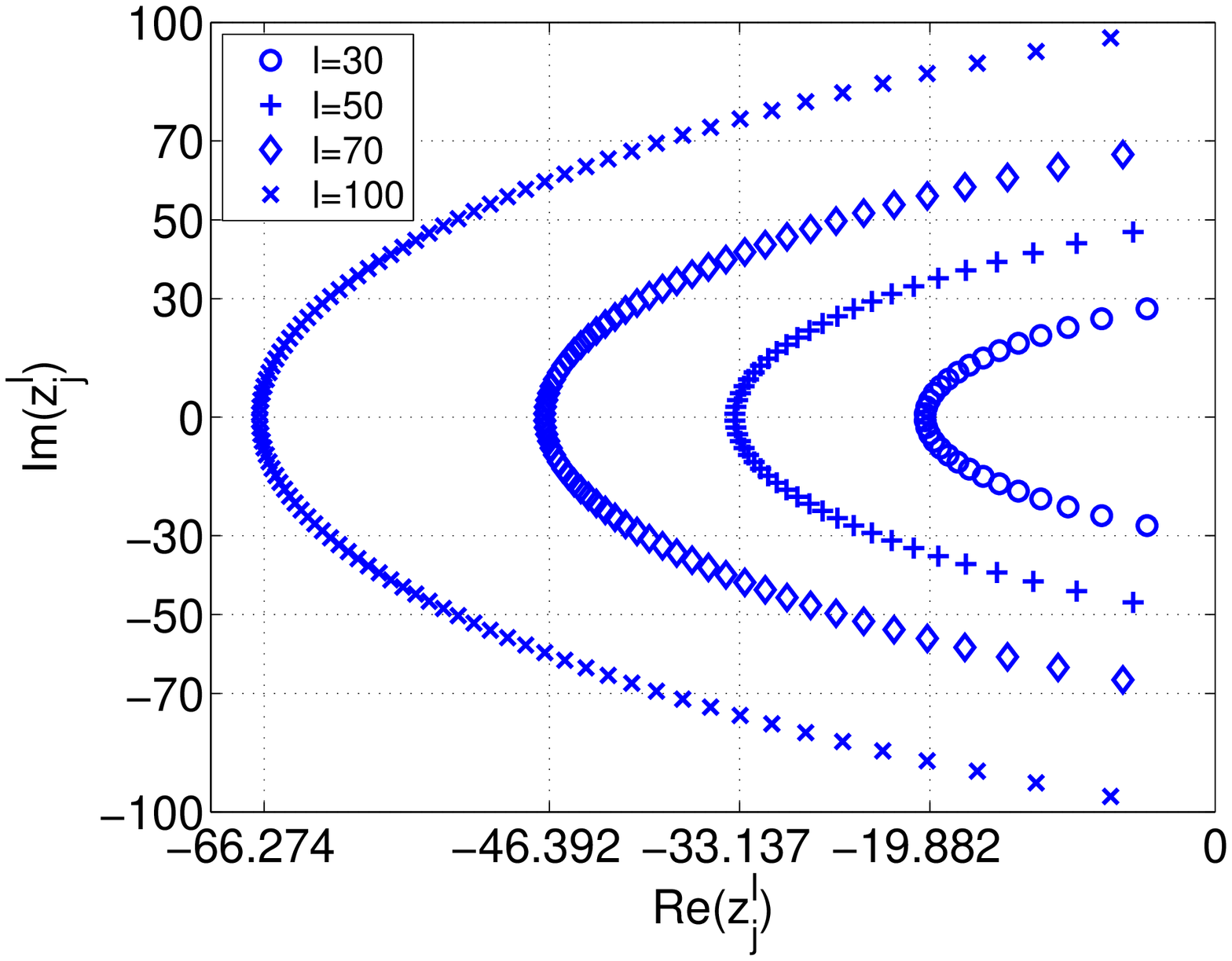}}
\end{minipage}}
\subfigure{
\begin{minipage}[t]{0.44\textwidth}
\centering
\rotatebox[origin=cc]{-0}{\includegraphics[width=0.75\textwidth,height=0.7\textwidth]{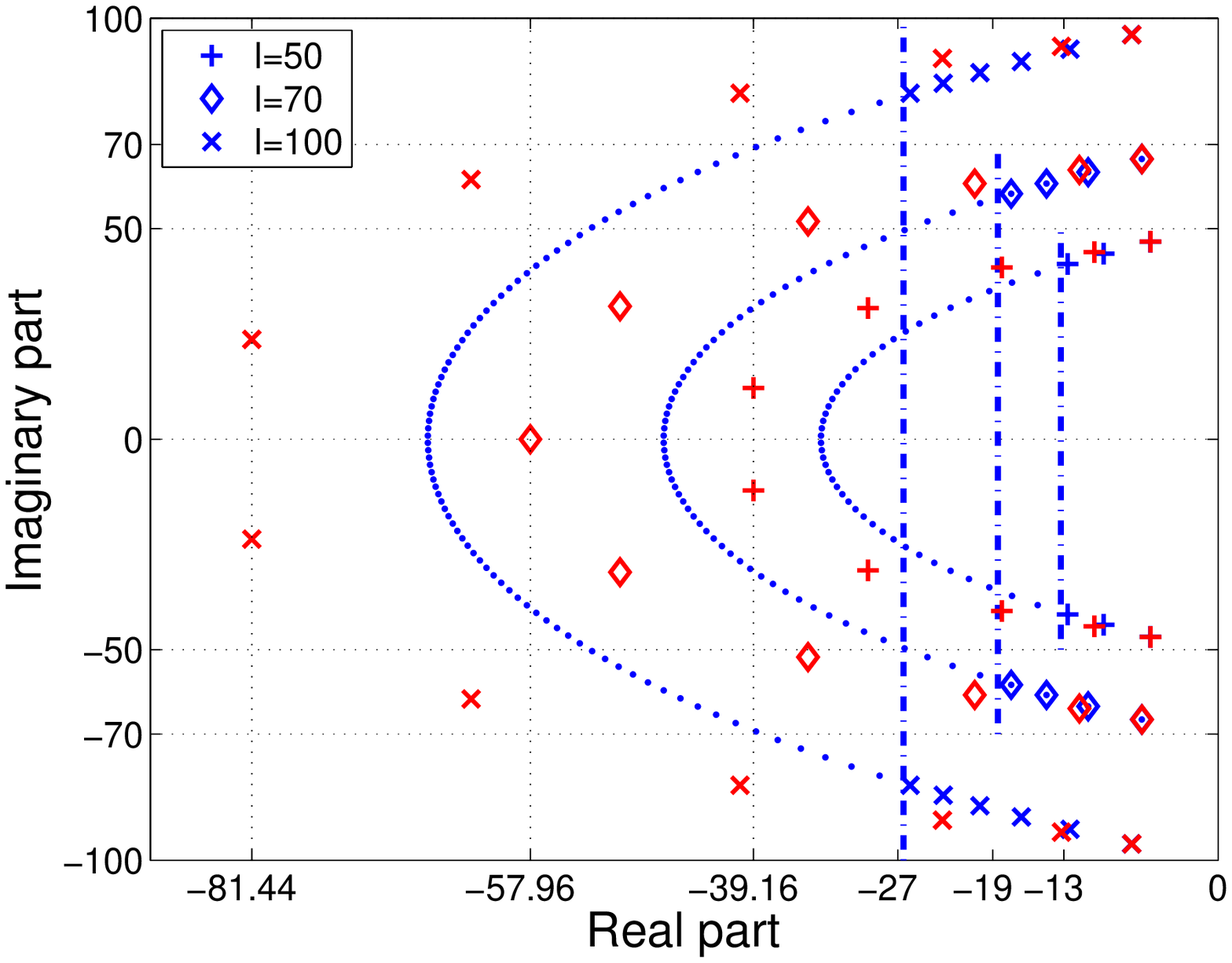}}
\end{minipage}}
\vspace*{-10pt} \caption{\small Left:  distributions of the zeros of
$K_{l+1/2}(z).$ Right:  distributions of poles in \cite{Alpert}
(red color), and used zeros (blue color), which lies on the right of
the  vertical dashdot line $-\beta la$ with
$\beta=0.4.$}\label{FigKzero}
\end{figure}

\noindent{\bf Remark 1.}~  We find from \cite{watson} that  (see Figure \ref{FigKzero} (left)): (i)   $K_{l+1/2}$  has exactly $l$ zeros,  which appear in conjugate pairs
and lie in the left-half of $z$-plane; and (ii) the zeros approximately sit along the boundary of an
eye-shaped domain  that  intersects the imaginary axis
approximately at $\pm {\rm i }\,l,$ and the negative real axis at $-la,$ where $a\approx 0.66274.$  We point out that a practical algorithm in \cite{LiJing2006} could be used to find the zeros  of $K_{l+1/2}(z)$  for any $l\le 1000$ accurately in negligible
time. \hfill  \rule{2mm}{2mm}

Armed with the explicit formulas \eqref{kernela}-\eqref{kernel}, we can compute the NRBKs at any time.
Observe that as the real part of $z_j^l$ is negative,  $e^{\frac{c}{b} z_j^lt}$ becomes exponentially small when  $z_j^l$
is far away from the imaginary axis and $t$ is slightly large.  This motivates us to drop many insignificant zeros by using
the algorithm described in \cite{Wang2Zhao12} (see Figure \ref{FigKzero} (right)).

\vskip 4pt
\noindent{\bf C.~ Fast algorithms for temporal convolution.}
\vskip 4pt

Observe from  \eqref{kernela}-\eqref{kernel} that  the time variable  $t$ only appears in the   exponentials.   This allows for
a fast  recursive temporal convolution as  shown in   \cite{Alpert}.
More precisely,  given a generic function  $g(t)$,  we define
 \begin{equation}\label{recurdeltab}
f(t;r):=e^{\frac{c}{b}rt}\ast g(t)=\int_0^t e^{\frac{c}{b}r(t-\tau)} g(\tau)\, d\tau,
\end{equation}
and find
 \begin{equation}\label{recurdeltabC}
 f(t+\Delta t; r)=e^{\frac{c}{b}r\Delta t} f(t; r)+\int_{t}^{t+\Delta t}e^{\frac{c}{b}r(t+\Delta t-\tau)} g(\tau)\, d\tau,
 \end{equation}
 where $\Delta t$ is the time step size.   We see that at each time step, the computation narrows down to computing the integral of the current interval $[t, t+\Delta t].$ This essentially eliminates the burden of history dependence induced by  the temporal  convolution.

 Using this notion, we deduce from \eqref{kernela}-\eqref{recurdeltab} that
\begin{align}
& [\sigma_l\ast g](t)=\frac c
{b} \sum_{j=1}^{l}z_j^l \int_0^t e^{\frac{c}{b}z_j^l(t-\tau)}g(\tau)d\tau =\frac
c {b} \sum_{j=1}^{l}z_j^l f(t;z_j^l),\label{recurdelta}\\
&[\omega_l\ast g](t)=\frac c
{b} \sum_{j=1}^{l}(z_j^l)^2 \int_0^t
e^{\frac{c}{b}z_j^l(t-\tau)}g(\tau)d\tau+g(t)\sum_{j=1}^l z_j^l=\frac c {b} \sum_{j=1}^{l}(z_j^l)^2 f(t;z_j^l)+g(t)\sum_{j=1}^l
z_j^l.\label{recurdeltassd}
\end{align}
Given $g$ at grids for time discretization,   we only need to store $\{f(t;z_j^l)\}_{j=1}^l$  for previous steps to compute
the convolutions at current time.  It is optimal for storage requirement.

 \vskip 4pt
\noindent{\bf D.~ Summary of the algorithm.}
\vskip 4pt

For clarity, we summarize the whole algorithm and outline two strategies to further reduce the complexity related to large $l.$
The algorithm is intended to compute ${\mathcal T}_b[\bs E](t)$ in  \eqref{newEtM} at $t=t_n=n\Delta t, n\ge 0,$ and on the colatitude-longitude grids adopted by the Spherepack \cite{adams1999spherepack} from $\bs E$ on the same grids.
\vskip 4pt
\noindent-----------------------------------------------------------------------------------------------------------------------------
\centerline{\bf Algorithm for computing  ${\mathcal T}_b[\bs E]$ in  \eqref{newEtM}}
\vskip 4pt
\begin{itemize}
\item[] Step 1.~  Use the Spherepack to compute $\big\{{E}_{lm}^{r},{E}_{lm}^{(1)}\big\}$ from  $\bs E.$
\item[] Step 2.~ Compute the zeros $\{z_j^l\}$ for  $l\ge 1.$
\item[] Step 3.~ Compute $\sigma_l \ast {E}_{lm}^{(2)}$ and  $\omega_l\ast {E}_{lm}^{r}$ via
\eqref{recurdelta}-\eqref{recurdeltassd}.
\item[] Step 4.~  Use the Spherepack to compute ${\mathcal T}_b[\bs E]$ via \eqref{newEtM}.
 \end{itemize}
 \vspace*{-8pt}
\noindent-----------------------------------------------------------------------------------------------------------------------------

It is clear that the number of zeros to be used is determined by the truncation of the expansion \eqref{Eexpansion}.  If $l$ is large,
we can adopt the strategies (i) dropping insignificant zeros (cf. \cite{Wang2Zhao12}) and (ii) compression algorithm (cf.
\cite{Alpert}) to reduce the complexity in Step 3.  Here, we outline the main idea.
\begin{itemize}
\item[(i).] Since   $e^{\frac{c}{b} z_j^lt}$ becomes exponentially small when  $z_j^l$
is far away from the imaginary axis and $t$ is slightly large, for some $t_0^l>0,$   we modify  the  NRBKs $\sigma_l$ and $\omega_l$ in \eqref{kernela}-\eqref{kernel} as follows
\begin{equation}\label{modifbc}
\tilde \sigma_l(t)=\frac c b\sum_{j\in  \Upsilon_l^\beta} z_j^l e^{\frac c bz_j^l t},\quad
\tilde \omega_l(t)=\frac c b \sum_{j\in  \Upsilon_l^\beta} (z_j^l)^2 e^{\frac{c}{b} z_j^lt}+\delta(t)\sum_{j=1}^l z_j^l,\;\; t>t_0^l,
\end{equation}
where  $  \Upsilon_l^\beta=\big\{ z_j^l\, :\,  {\rm Re} (z_j^l) \ge -\beta l a   \big\}$ with  $  a\approx 0.66742$
and  $\beta\in (0,1)$ (\,$\beta$ tunes the number of used zeros).  We plot in Figure
\ref{FigKzero} (right) the used zeros for $\beta=0.4$. This can reduce the zeros from $100$ to $10$
(see the marker `$\times$' on the right of vertical dashdot line) and leads to quite accurate approximation.
We refer to the analysis in \cite{Wang2Zhao12} on how to  adjust $\beta$ and $t_0^l$ to achieve a good accuracy.

\item[(ii).] Alpert et al. \cite{Alpert} proposed a compression technique by a rational approximation of the NRBK in $s$-domain, which required  to solve a nonlinear least square problem.  This led to the approximate poles $\{\tilde{z}_j^d\}_{j=1}^d$ with $d\ll l$  (see Figure \ref{FigKzero} (right,  marked by `$\times$') with a reduction  from $l=100$ to $d=12$ and error tolerance $10^{-8}$).
  Correspondingly,   $\sigma_l$ and $\omega_l$ could be  approximated by
\begin{equation}\label{kernelalpert}
\hat \sigma_{l}(t)=\frac{c} b \sum_{j=1}^{d}\alpha_j^d e^{\frac{c}{b} \tilde z_j^dt},
\;\;
 \hat \omega_l(t)=\frac c b \sum_{j=1}^d \alpha_j^d\tilde z_j^d e^{\frac{c}{b} \tilde z_j^dt}+\delta(t)\sum_{j=1}^d \alpha_j^d,\quad t>0,
 \end{equation}
 where $\{\alpha_j^d\}_{j=1}^d$ are the coefficients occurring in the rational partial fraction:  $\alpha_j^d/(s-c\tilde z_j^d/b).$
Some samples of $\{\alpha_j^d, \tilde{z}_j^d\}_{j=1}^d$ are available from the website: http://faculty.smu.edu/thagstrom/sph6.txt.
 \end{itemize}

\vskip 12pt
\centerline{\bf 3.~ Numerical Results and  Discussions}
\vskip 12pt
In this section, we provide some numerical examples to show the accuracy  of computing the NRBC.
We also test a spectral-Galerkin with (second-order) Newmark's  time integration for  Maxwell's equations in a spherical shell
$\{a<|\bs x|<b\}$, where the NRBC is set at the outer spherical surface $r=b$.

 In the following tests, we generate the exact solution through the field: $(\bs  {\hat  x}\times \bs E)|_{r=a}=\bs g$, and with homogeneous initial data and source term. More precisely, we take
$${\bs g}(\theta,\phi,t)=\sum_{l=1}^{\infty}\sum_{m=-l}^l\bigg[
\sin^{6}(6t)\, g_{l,1}^m\,\nabla_SY_l^m+ \bigg(\mathcal
L^{-1}\bigg[\frac{c}{sa}+\frac{k'_l(sa/c)}{k_l(sa/c)}\bigg](t)\ast\sin^{8}(4t)\bigg)\,g_{l,2}^m\,{\bs
T}_l^m\bigg],$$
where  $\{g_{l,1}^m, g_{l,2}^m\}$ are the expansion coefficients in  terms of $\{\nabla_SY_l^m,  \vt\},$  of the
 the function
$${\tilde {\bs g}}(\theta,\phi)=b\bigg(\,\Big(\cos\big(x_1^2x_2x_3\big)+2\Big)^{\frac{5}{2}},
x_3^2\Big(\sin\big(x_1x_2\big)+1\Big)^{\frac{3}{2}},x_1\Big(0.5\cos\big(x_2x_3\big)+1\Big)^{\frac{7}{2}}\,\bigg).$$
Note that $\bs x=(x_1, x_2, x_3)$ (with $|\bs x|=1$) is the corresponding  Cartesian coordinates.
Here, we compute $\{g_{l,1}^m, g_{l,2}^m\}$ accurately by using the Spherepack \cite{adams1999spherepack}.

 We first test the accuracy of the algorithm for the NRBC \eqref{NRBCold}.
Let ${\bs E}^{\bs N}$ be the truncated exact solution $\bs E$ (the included modes are  $1\le l\le N_\theta; |m|\le l$).
 Define the error: 
\begin{equation}\label{NRBCerr}
e(b,t)=\Big\|\Big(\partial_t {\bs E}^{\bs
N}_T-c\,\bs {\hat x}\times (\nabla \wedge {\bs E}^{\bs
N})\Big)-{\mathcal T}_b[{\bs E}^{\bs
N}]\Big\|_{N_\theta,N_\phi},
\end{equation}
where  $\|\cdot\|_{N_\theta,N_\phi}$ denotes the discrete $L^2$-norm associated with  $N_\theta\times N_\phi$ colatitude and longitude grids. Here, we take   $a=2, c=5, N_\theta=40$ and  $N_\phi=2N_\theta$. We aim  to test the accuracy for computing
${\mathcal T}_b[{\bs E}^{\bs N}],$ so the differentiations in $t$ and curl are calculated analytically.
 In Table \ref{tb1new}, we tabulate  $e(b,t)$ for different $t$ and $b$ (note: the magnitude of  $\bs E$ is actually between $1$ and $20$, so the waves cross the artificial boundary).  We see that in all cases, the computation of the NRBC is very accurate.

\vspace*{-8pt}
\begin{table}[!ht]
{\footnotesize
\begin{center}
 \caption{\small The error $e(b,t)$ for different $t$ and $b$}\label{tb1new}
 \vskip 6pt
\begin{tabular}{c c c c c}
\hline\hline
\multicolumn{1}{c}{{$t$}}&
\multicolumn{1}{c}{{$b=3$}}&\multicolumn{1}{c}{{$b=5$}}&
\multicolumn{1}{c}{{$b=5$}}&
\multicolumn{1}{c}{{$b=6$}}
\\
    \cline{1-5}
  1.0  &   7.73246E-14     &      2.67652E-14            &  1.64591E-14   &  1.10549E-14\\
  2.0  &  7.01963E-14   &  4.43072E-14   & 4.30850E-14  &  9.58167E-14  \\
4.0  & 1.30672E-13  &  7.07636E-14   &       8.15754E-14                 &  9.56271E-14  \\
10.0 & 3.35293E-13   &       1.87063E-13                  &  2.32403E-13 &   2.74720E-13 \\
\hline\hline
\end{tabular}
\end{center}
}
\end{table}

Next, we set the NRBC as the boundary condition and solve the Maxwell's equation in curl-curl formulation with homogeneous initial conditions and free source in a spherical shell. In this case, we can expand the interior electric field in VSHs, and reduce the problem
to a sequence of equations in radial direction. Then we solve the systems by using the spectral-Galerkin method in space and
the second-order Newmark scheme in time.  Moreover, we use the  Richardson extrapolation  to improve
the time discretization to fourth-order. We refer to \cite{Wang2Zhao12} for similar idea for the acoustic wave equations, and report the details in the extended version.

Under the same setting of the reference solution and other data, we provide in  Table \ref{tb2new},
 the discrete $L^2$-norm errors (in space with sufficient resolution):
  $\text {Err(}t\text{)}$ (Newmark scheme), $\text {Err$^{\text R}$(}t\text{)}$ (Richardson extrapolation),  and the convergence order in time  at different time $t$. As expected, we observe  the second-order convergence for the Newmark scheme  and the fourth-order convergence for the extrapolation.

\begin{table}[!ht]
{\footnotesize
\begin{center}
 \caption{\small Convergence of the Newmark scheme and Richardson extrapolation.} \vspace*{-4pt}
\begin{tabular}{c c c c c c c c c c c c }
\hline\hline
\multicolumn{1}{c}{{\raisebox{-0.4ex}[0pt]{$t$}}}&
\multicolumn{1}{c}{{\raisebox{-0.4ex}[0pt]{$\Delta t$}}}&
\multicolumn{1}{c}{{\raisebox{-0.4ex}[0pt]{Err}}}&
\multicolumn{1}{c}{{\raisebox{-0.4ex}[0pt]{order}}} &
 \multicolumn{1}{c}{{\raisebox{-0.6ex}[0pt]{Err$^{\text R}$ }}}&
\multicolumn{1}{c}{{\raisebox{-0.4ex}[0pt]{order}}}&\multicolumn{1}{c}{{\raisebox{-0.4ex}[0pt]{$t$}}}&
\multicolumn{1}{c}{{\raisebox{-0.4ex}[0pt]{$\Delta t$}}}&
\multicolumn{1}{c}{{\raisebox{-0.4ex}[0pt]{Err}}}&
\multicolumn{1}{c}{{\raisebox{-0.4ex}[0pt]{order}}} &
 \multicolumn{1}{c}{{\raisebox{-0.6ex}[0pt]{Err$^{\text R}$ }}}&
\multicolumn{1}{c}{{\raisebox{-0.4ex}[0pt]{order}}}
\\
\cline{1-12}
         & 5.00e-3  & 2.3090E-2 &             & 2.0719E-5 &     & & 5.00e-3 & 1.6912E-2 &             & 2.8491E-5 &           \\
         & 2.50e-3  & 5.7684E-3 &    2.001        & 1.2870E-6 &    4.009  &  & 2.50e-3 &  4.2193E-3 & 2.003 &  1.7754E-6&        4.004      \\
  {\raisebox{1.4ex}[0pt]{0.5}}         & 1.25e-3  & 1.4419E-3   & 2.000  &8.0317E-8 & 4.002  &  {\raisebox{1.4ex}[0pt]{1.5}}& 1.25e-3  &  1.0543E-3   &  2.000 &   1.1088E-7  &   4.001 \\
         & 1.00e-3  & 9.2276E-4 &     2.000        & 3.2912E-8 &    3.998 & & 1.00e-3  &  6.7470E-4 &    2.001        & 4.5411E-8 &    4.009    \\
    \cline{1-12}
         & 5.00e-3  & 2.1041E-2 &             & 2.6540E-5 &  &  & 5.00e-3  & 2.2117E-2 &             &  2.3189E-5 &        \\
         & 2.50e-3  &5.2591E-3&    2.000        & 1.6500E-6 &   4.008  &  & 2.50e-3 & 5.5142E-3 & 2.004  &  1.4467E-6 &      4.003   \\
   {\raisebox{1.4ex}[0pt]{1.0}}        & 1.25e-3  &  1.3147E-3  & 2.000 & 1.0299E-7 & 4.001 & {\raisebox{1.4ex}[0pt]{2.0}} & 1.25e-3  & 1.3776E-3  &  2.001 &  9.0376E-8  &    4.001   \\
         & 1.00e-3  & 8.4140E-4 &    2.000        & 4.2178E-8&    4.009   &  & 1.00e-3  & 8.8159E-4 &    2.000        &  3.7016E-8 &    4.000 \\
\hline\hline
\end{tabular}\label{tb2new}
\end{center}
}
\end{table}

We have proposed in this paper  an efficient algorithm for computing the spherical NRBC for the Maxwell's equations.
This tool can be integrated well with various interior solvers in bounded domain for simulating scattering problems in many
situations. We will report the works along this line in the forthcoming papers.

\baselineskip 12pt
\small


\begin{thebibliography}{10}

\bibitem{adams1999spherepack}
J.C. Adams and P.N. Swarztrauber.
\newblock Spherepack 3.0: A model development facility.
\newblock {\em Monthly Weather Review}, 127(8):1872--1878, 1999.

\bibitem{Alpert}
B.~Alpert, L.~Greengard, and T.~Hagstrom.
\newblock Rapid evaluation of nonreflecting boundary kernels for time-domain
  wave propagation.
\newblock {\em SIAM J. Numer. Anal.}, 37(4):1138--1164 (electronic), 2000.

\bibitem{Bere94}
J.P. Berenger.
\newblock A perfectly matched layer for the absorption of electromagnetic
  waves.
\newblock {\em J. Comput. Phys.}, 114(2):185--200, 1994.

\bibitem{Chen08}
Z.M. Chen and J.-C. N{\'e}d{\'e}lec.
\newblock On {M}axwell equations with the transparent boundary condition.
\newblock {\em J. Comput. Math.}, 26(3):284--296, 2008.

\bibitem{CISB91}
R.D. Ciskowski and C.A. Brebbia.
\newblock {\em Boundary Element Methods in Acoutics}.
\newblock Kluwer Academic Publishers, 1991.


\bibitem{Hagstrom99}
T.~Hagstrom.
\newblock Radiation boundary conditions for the numerical simulation of waves.
\newblock In {\em Acta numerica, 1999}, volume~8 of {\em Acta Numer.}, 
  47--106. Cambridge Univ. Press, Cambridge, 1999.

\bibitem{Hagstrom07}
T.~Hagstrom and S.~Lau.
\newblock Radiation boundary conditions for {M}axwell's equations: a review of
  accurate time-domain formulations.
\newblock {\em J. Comput. Math.}, 25(3):305--336, 2007.

\bibitem{Kel.G89}
J.B. Keller and D.~Givoli.
\newblock Exact nonreflecting boundary conditions.
\newblock {\em J. Comput. Phys.}, 82(1):172--192, 1989.

\bibitem{lee1995whitney}
J.F. Lee and Z.~Sacks.
\newblock Whitney elements time domain (WETD) methods.
\newblock {\em IEEE Transactions on Magnetics}, 31(3):1325--1329, 1995.

\bibitem{LiJing2006}
J.R. Li.
\newblock Low order approximation of the spherical nonreflecting boundary
  kernel for the wave equation.
\newblock {\em Linear Algebra Appl.}, 415(2-3):455--468, 2006.

\bibitem{lin2009hybrid}
Y.~Lin, J.H. Lee, J.~Liu, M.~Chai, J.A. Mix, and Q.H. Liu.
\newblock {A hybrid SIM-SEM method for 3-D electromagnetic scattering
  problems}.
\newblock {\em IEEE Transactions on Antennas and Propagation},
  57(11):3655--3663, 2009.

\bibitem{liu1998pstd}
Q.H. Liu.
\newblock The {PSTD} algorithm: A time-domain method requiring only two cells
  per wavelength.
\newblock {\em Microwave and Optical Technology Letters}, 15(3):158--165, 1998.

\bibitem{Lubistian2002}
C.~Lubich and A.~Sch{\"a}dle.
\newblock Fast convolution for nonreflecting boundary conditions.
\newblock {\em SIAM J. Sci. Comput.}, 24(1):161--182 (electronic), 2002.

\bibitem{Morse53}
P.M. Morse and H.~Feshbach.
\newblock {\em Methods of Theoretical Physics. 2 volumes}.
\newblock McGraw-Hill Book Co., Inc., New York, 1953.

\bibitem{Nedelec}
J.C. N{\'e}d{\'e}lec.
\newblock {\em Acoustic and Electromagnetic Equations}, volume 144 of {\em
  Applied Mathematical Sciences}.
\newblock Springer-Verlag, New York, 2001.
\newblock Integral representations for harmonic problems.

\bibitem{Nich.Shen2009}
D.~Nicholls and J.~Shen.
\newblock A rigorous numerical analysis of the transformed field expansion
  method.
\newblock {\em SIAM J. Numer. Anal.}, 47(4):2708--2734, 2009.

\bibitem{Sofronov1992}
I.L. Sofronov.
\newblock Conditions for complete transparency on a sphere for a
  three-dimensional wave equation.
\newblock {\em Dokl. Akad. Nauk}, 326(6):953--957, 1992.

\bibitem{TafloveHag.05}
A.~Taflove and S.C. Hagness.
\newblock {\em Computational Electrodynamics: the Finite-Difference Time-Domain
  Method}.
\newblock Artech House Inc., Boston, MA, third edition, 2005.

\bibitem{Wang2Zhao12}
L.-L. Wang, B.~Wang, and X.D. Zhao.
\newblock Fast and accurate computation of time-domain acoustic scattering
  problems with exact nonreflecting boundary conditions.
\newblock {\em SIAM J. Appl. Math.}, 72(6):1869--1898, 2012.

\bibitem{watson}
G.N. Watson.
\newblock {\em A Treatise of the Theory of Bessel Functions (second edition)}.
\newblock Cambridge University Press, Cambridge, UK, 1966.

\end{thebibliography}

\end{document}